\documentclass[journal]{IEEEtran}
\usepackage{cite}
\usepackage{amsmath,amssymb,amsfonts}
\usepackage{algorithmic}
\usepackage{graphicx}
\usepackage{textcomp}
\usepackage{xcolor}

\usepackage{atbegshi}
\AtBeginDocument{\AtBeginShipoutNext{\AtBeginShipoutDiscard}}

\begin{document}
\title{Impact of Orientation Misalignments on Black Phosphorus Ultrascaled Field-effect Transistors}

\author{Cedric Klinkert, Sara Fiore, Jonathan
    Backmann, Youseung Lee, and Mathieu Luisier\\
Integrated System Laboratory, ETH Zurich, Switzerland (e-mail: cedrick@iis.ee.ethz.ch)}.

\maketitle

\begin{abstract}
Two-dimensional materials with strong bandstructure anisotropy such as
black phosphorus {\color{black}(BP)} have been identified as attractive
candidates for logic application due to their potential high carrier
velocity and large density-of-states. However, perfectly aligning the
source-to-drain axis with the desired crystal orientation remains an
experimental challenge. In this paper, we use an advanced quantum
transport approach from first-principle to shed light on the influence
of orientation misalignments on the performance of {\color{black} BP}-based
field-effect transistors. Both $n$-and $p$-type configurations are
investigated for six alignment angles{\color{black}, in the ballistic
  limit of transport and in the presence of electron-phonon and
  charged impurity scattering}. It is found that up to
deviations of $50^{\circ}$ from the optimal angle, the ON-state
current only decreases by $30\%$. This behavior is explained by
considering a single bandstructure parameter, the effective mass along
transport direction. 
\end{abstract}



\section{Introduction}
\IEEEPARstart{2D}{Materials} have received a wide attention from the
scientific community ever since the discovery of graphene in 2004
\cite{GrapheneDiscov}. The necessity of a band gap for logic
applications ultimately led experimental efforts towards other 2D
materials such as transition metal dichalcogenides (TMDs)
e.g. MoS$_2$ \cite{MoS2Transistor} {\color{black} black phosphorus (BP)
  \cite{Li2014_Rev2_1,Liu_Rev2_2,Xia2014_Rev2_3}. The latter material
  emerged} as a serious contender due to its anisotropic bandstructure 
that simultaneously provides high carrier velocities and large
density-of-states (DOS) {\color{black}
  \cite{LiuP4_Rev1_2,HartipourP4withHfo2gate_Rev1_3,Qiao2014_Rev2_5},
  two properties that have been leveraged to design various types of
  more-than-Moore logic switches
  \cite{Wu2019_Rev2_10,Ilatikhameneh2016_Rev4_1}}.

Recent theoretical studies \cite{2D_database_Mounet, Klinkert} suggest
that many other 2D materials with strong anisotropic conduction (CB)
and/or valence band (VB) exist, e.g. $Ag_2N_6$, $S_6Te_4Zr_2$ or
$P_8Si4$. In their mono-layer form, when used as the channel material
of ultra-scaled field-effect transistors, such compounds might exhibit
almost ideal electrostatic characteristics as well as very high
ON-state currents, if the transport direction is aligned with the
proper crystal axes, i.e. the one with the lowest effective
mass. However, from an experimental point of view, obtaining such
perfect alignments is extremely challenging. If transport occurs along
the crystal axis with the largest effective mass, {\color{black} device
  simulation predicts severe performance degradations for BP
  \cite{Liu2014_Rev2_6,Lam2014_Rev2_7,Quhe2018_Rev2_9}. Most studies
  focused on the best- and worst-case scenarios, very few on what
  happens in between \cite{Wan2016_Rev3_1}}.

Here, we therefore study the impact of orientation misalignments on
the ON-state current of ultra-scaled {\color{black} BP}-FETs in
their $n$- and $p$-type configuration. An advanced quantum transport
solver combining density-functional theory (DFT) and the
Non-equilibrium Green's Function formalism (NEGF) is employed for that
purpose. After introducing this simulation approach in Section
\MakeUppercase{\romannumeral 2}, it is revealed in Sections
\MakeUppercase{\romannumeral 3} that orientation misalignments up to
$50^{\circ}$ from the ideal case do not significantly alter the
ON-state current of {\color{black} BP}-FETs, with minor performance
loss up to a misalignment angle of $20^{\circ}$. {\color{black} This
  behavior, which occurs both in the ballistic limit of transport and
  in the presence of electron/hole-phonon and charged impurity
  scattering, is explained based on the calculation of angle-dependent
  effective masses.} 


\begin{figure}[h]
    \centering
    \includegraphics[width=0.5\textwidth]{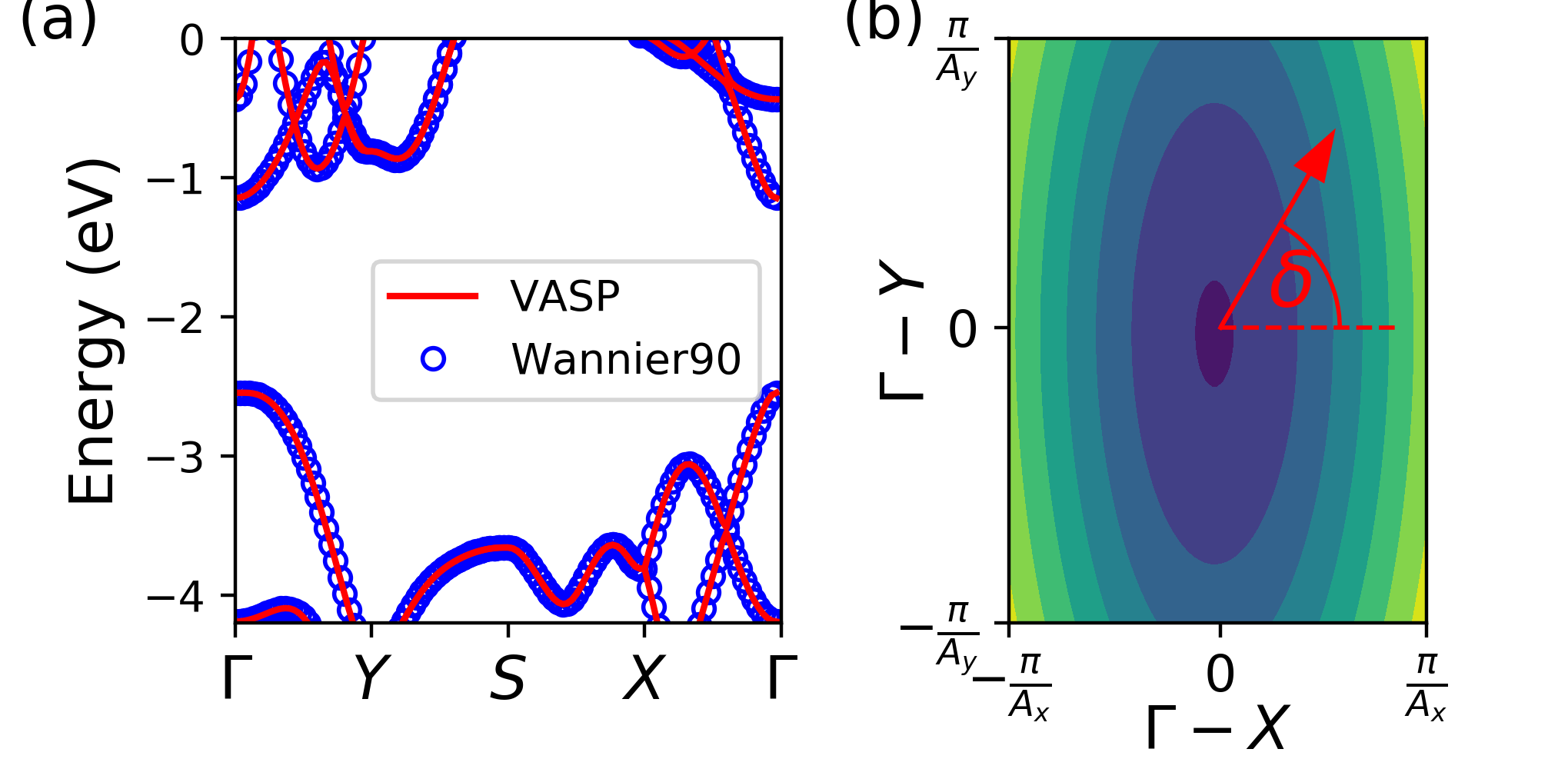}
    \caption{
    (a) Electronic bandstructure of a {\color{black} BP} monolayer as
      calculated with the VASP DFT code (red lines) and after
      transformation into a basis of maximally localized Wannier
      functions (blue circles). 
    (b) Energy contour-plot of the lowest conduction band of
      {\color{black} BP}. The red arrow indicates a transport direction
      with a misalignment angle $\delta$  with respect to the
      direction having the smallest transport effective mass.} 
    \label{fig:fig1}
\end{figure}

\section{Approach}
The simulation results reported in this paper were obtained with an
atomistic quantum transport solver \cite{OMEN_Mathieu} based on the
NEGF equations {\color{black} following the same methodology as in
  \cite{Klinkert}}. An \emph{ab initio} scheme is employed to capture
the electronic band structure of {\color{black} BP} and to create a
suitable Hamiltonian matrix for the device geometry of interest. This
step is achieved by converting the electronic structure received from
a DFT package such as VASP \cite{VASP_code} from a delocalized
plane-wave basis to a set of maximally localized Wannier functions
(MLWF) using the Wannier90 code \cite{Wannier_code}. {\color{black} 
  The Hamiltonian of any unit cell can now be constructed 
  by identifying the Wannier interaction for each atomic bond \cite{stieger}.
  To minimize the computational burden, transport is 
  evaluated along directions leading to reasonably small orthorhombic
  unit cells}.  

As exchange correlation functional the general gradient approximation
of Perdew Burke Ernzerhof \cite{PBE} is used. The PBE band gap of
{\color{black} BP},  $E_{g,PBE}=0.91\;eV$, is raised to the value
obtained by the GLLBsc functional \cite{GLLBsc}, $E_{g,GLLBsc} =
1.40\;eV$ to avoid artificial source-to-drain tunneling in the
OFF-state. This increase is justified by the systematic
underestimation of the semiconductor band gaps by the  PBE
functional. The resulting bandstructure of {\color{black} BP}, before
and after the MLWF transformation is shown in Fig. \ref{fig:fig1}(a)
demonstrating the accuracy of the approach. A contour plot of the
lowest conduction band is depicted in Fig. \ref{fig:fig1}(b). The
optimal performance is obtained when the transport axis is aligned
with the $\Gamma-X$ axis, while $\delta$ indicates the misalignment
angle.

{\color{black} As scattering is known to influence electron and hole
  transport in BP transistors \cite{LiuP4_Rev1_2,Gaddemane2018_Rev2_8},
  the NEGF model of \cite{lee} is adopted to account for these
  effects through dedicated self-energies. It combines charged
  impurity scattering (CIS), as defined in \cite{ong}, and
  electron/hole-phonon interactions with \textit{ab initio} inputs
  only, as described in \cite{szabo}. Both the derivatives of the
  Hamiltonian matrix, which couple the electron/hole and phonon
  populations, and the phonon modes/frequencies are computed at the
  DFT level with VASP.}


\begin{figure}[h]
    \centering
    \includegraphics[width=0.5\textwidth]{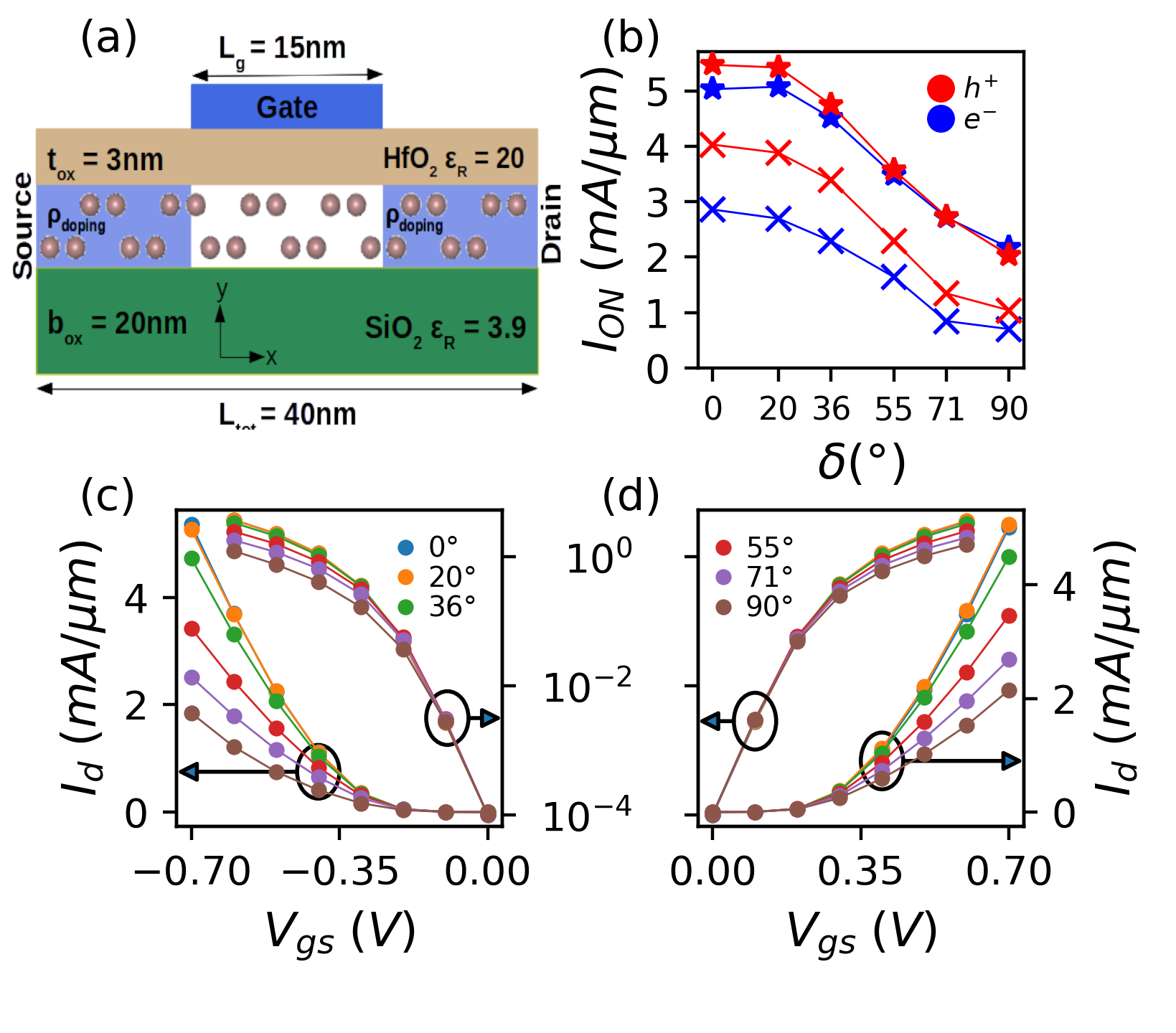}
    \caption{(a) Schematic view of the simulated single-gate FET with
      a {\color{black} BP} mono-layer as channel. 
             (b) ON-state currents of all simulated transistors as a
      function of the misalignment angle $\delta${\color{black}, with
        (lines with crosses) and without (lines with stars) CIS and
        electron/hole-phonon scattering.} 
             (c) {\color{black} Ballistic} transfer characteristics
      $I_d-V_{gs}$ at $V_{ds} = 0.7\; V$ for $p$-type {\color{black} BP}
      FETs as in (a) at different misalignment angles $\delta$.
             (d) Same as (c), but for $n$-type FETs.}
    \label{fig:fig2}
\end{figure}
  
\section{Results}
As an example to illustrate the influence of orientation
misalignments, we designed single-gate {\color{black} BP} FETs loosely
inspired by the specifications of the International Roadmap for
Devices and Systems, as shown in Fig. \ref{fig:fig2}(a). A gate length
$L_g$=15 nm {\color{black} was chosen to minimize source-to-drain
  leakages, which are not the subject of this study}. All devices have
a total length $L_{tot} \approx 40\;nm$. The top oxide has an
equivalent oxide thickness $EOT = 0.6\;nm$ that ensures an excellent
electrostatic control. Source and drain extensions are doped with a
donor/acceptor carrier concentration of $5 \times 10^{13}\;cm^{-2}$.
We found that such high values do not artificially boost the ON-current
\cite{Klinkert}. The OFF-state current if fixed to $0.1\;\mu A / \mu
m$.

All simulations were carried out at a supply voltage $V_{ds}=0.7\;V$
and at room temperature. {\color{black} In Fig. \ref{fig:fig2}(a) only
  the P atoms belonging to the 2-D channels enter the NEGF
  framework. The other layers are accounted for through Poisson's
  equations. The source and drain contact resistances are assumed to
  be negligibly small. Realistic top or side contacts might influence
  the dependence of the current on the transport direction, but adding
  them is not compatible with the chosen MLWF approach where the same
  unit cell is repeated to construct the device channel. Similarly,
  other potentially relevant effects such as the formation of
  direction-dependent crystal or edge defects are not considered here
  because their treatment significantly increases the computational
  burden.}

Six different transport directions with a misalignment angle $\delta
\in \left \lbrace 0^{\circ},\; 20^{\circ},\; 36^{\circ},\;
55^{\circ},\; 71^{\circ},\; 90^{\circ}\right \rbrace$ w.r.t. the
optimal direction are considered. The ON-state currents of the
different $n$-and $p$-type transistors are reported in
Fig. \ref{fig:fig2}(b) as a function of $\delta$, {\color{black}
  with and without CIS and electron/hole-phonon scattering. For
  simplicity, the electrostatic potentials without scattering are used
  in both types of calculation. Finally, to enable rapid convergence
  of the NEGF calculations, small impurity concentrations of
  $N_{imp}$=2.5e11 cm$^{-2}$ are assumed in all cases.

The corresponding $I_d-V_{gs}$ ballistic transfer characteristics are
given in Figs. \ref{fig:fig2}(c) and (d). It can be observed that the
ON-state currents do not vary much for $\delta \leq 20^{\circ}$, even
in the presence of scattering.} Moreover, the current reduction, as
compared to the highest possible value is limited to $30\%$ for
misalignment $\delta<50^{\circ}$. The current degradation then rapidly
increases to reach $60\%$ or more at $\delta=90^{\circ}$. This finding
demonstrates that the performance loss does not linearly depend on the
misalignment angle, thus allowing for a certain tolerance during the
fabrication process.

\begin{figure}[h]
    \centering
    \includegraphics[width=0.49\textwidth]{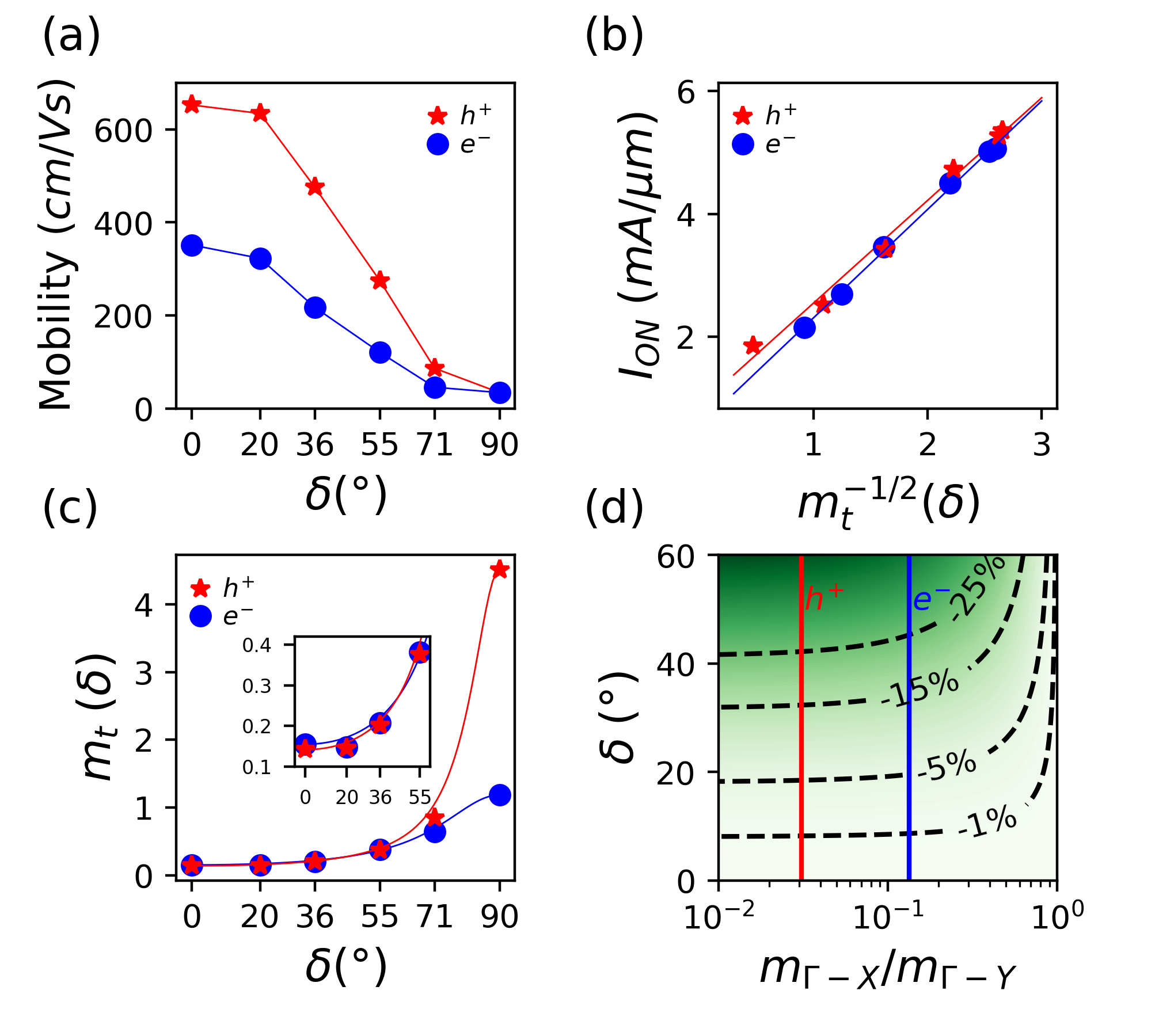}
    \caption{
    (a) {\color{black} Charged impurity and electron/hole-phonon limited
        low-field mobility of electrons and holes in BP as a function
        of the transport direction. An impurity concentration
        $N_{imp}$=2.5e11 cm$^{-2}$ is assumed.}
    (b) ON-state current from Fig. \ref{fig:fig2}(b) \emph{versus}
      $1/\sqrt{m_t\left(\delta\right)}$. A linear dependence between
      them is clearly visible. 
    (c) Transport effective mass $m_t\left(\delta\right)$ vs. $\delta$
      as obtained from Eq. \ref{dos_relation} and \ref{eq_current}
      (symbols) and as calculated with by Eq. \ref{eq_mt} (lines). 
    (d) Current reduction w.r.t. $\delta=0^{\circ}$ as a function of
      the $m_{\Gamma-X}/m_{\Gamma-Y}$ ratio and the alignment
      $\delta$, as computed with Eq. \ref{eq_current}. The dashed
      lines represent specific current reductions. The effective mass
      ratios of {\color{black} BP} are indicated by the vertical solid
      lines. 
    }
    \label{fig:fig3}
\end{figure}

{\color{black} The non-linear dependence of the ON-state current on the
  transport direction does not appear to be a consequence of CIS or
  electron/hole-phonon scattering, as confirmed by the corresponding
  electron and hole low-field mobility in Fig. \ref{fig:fig3}(a). This
  quantity, calculated with the ``dR/dL'' method \cite{rim}, exhibits
  a similar behavior as the ON-state current. It stays almost
  constant up to $20^{\circ}$, slightly decreases between $20^{\circ}$
  and $36^{\circ}$, before experiencing a faster reduction at greater
  angles. The same happens if the impurity concentration is increased
  to $N_{imp}$=1e12 cm$^{-2}$. These results tend to indicate that the
  non-linearity of the current is an intrinsic property of BP directly
  coming from its bandstructure.}

{\color{black} To reveal the role played by the BP bandstructure, we
  start by extracting} relevant physical quantities at the
top-of-the-barrier location (ToB) of the simulated FETs according to
\cite{Klinkert}. First, we observed that the charge at the ToB is
about the same, regardless of the misalignment angle, {\color{black}
  because} the DOS effective mass is orientation-independent. This
will be demonstrated later. An approximation for the gate capacitance
can be derived from the slope of the ToB charge density. A value $C_g$
$\approx 5.6\; \mu F / cm^2$ is found in all cases, close to the oxide
capacitance $C_{ox} \approx 5.9\; \mu F / cm^2$.

The ON-state current variations can be explained by approximating the
$E\left(\vec{k}\right)$ bandstructure dispersion of the lowest
conduction band or the highest valence band as a parabola with a
$\delta$-dependent inverse effective mass tensor $\left(1/m_{\delta}\right)$ 
\begin{align}
    E\left(\vec{k}\right) &= E_0 \pm \frac{\hbar^2}{2} \vec{k}^T \left(\frac{1}{m_{\delta}}\right) \vec{k}.
    \label{energy_relation}
\end{align}
The $+/-$ sign refers to the electron/hole case, the energy $E_0$
either to the CB minimum or VB maximum, and $\hbar$ to Plank's
reduced constant. In Eq. \ref{energy_relation} $1/m_{\delta}$ is
defined as 
\begin{equation}
    \frac{1}{m_{\delta}} = R_{\delta}^{T}\left(\frac{1}{m_{\delta=0^{\circ}}}\right) R_{\delta}\;
    , \quad R_{\delta} = 
    \begin{pmatrix}
    \cos\delta & \sin\delta\\
    -\sin\delta & \cos\delta\\
    \end{pmatrix}
    \label{rotation}
\end{equation}
where $R_{\delta}$ is a rotation matrix and $m_{\delta=0^{\circ}}^{-1}
= diag\left(m_{\Gamma-X}^{-1},m_{\Gamma-Y}^{-1} \right)$, the inverse
effective mass tensor for transport along the $\Gamma-X$ direction
with $m_{\Gamma-X} = 0.16m_0\;\left(0.14m_0 \right)$, and
$m_{\Gamma-Y} = 1.20m_0 \; \left(4.46m_0\right)$ for electrons
(holes). {\color{black} These effective masses were extracted as in
  \cite{Klinkert} for an electron/hole population of $\sim$1.5e13
  cm$^{-2}$}.  The transport effective mass $m_t\left(\delta\right)$
along any direction can be computed from Eq. \ref{rotation} as
{\color{black} \cite{Wu2019_Rev2_10}} 
\begin{equation}
    \frac{1}{m_{t}\left(\delta\right)} = \frac{\sin^2\delta}{m_{\Gamma-Y}} +  \frac{\cos^2\delta}{m_{\Gamma-X}} 
    \label{eq_mt}
\end{equation}
By combining Eq. \ref{energy_relation} and \ref{rotation} the 2D DOS can be written as
\begin{equation}
    DOS_{2D}\left(\epsilon\right) =
    \frac{\sqrt{m_{\Gamma-X}m_{\Gamma-Y}}}{\pi\hbar^2}H\left(\pm\left(\epsilon-E_0\right)\right)
    \label{dos_relation}
\end{equation}
Here, $H$ is the Heavy-side function. It can be seen that the DOS only
depends on the effective masses along the principle axes of the
ellipses describing the bands, not on $\delta$. This is why the gate
capacitance, and the electrostatics too, are the same for all
transport directions.

Next, an analytical expression is derived for the $\delta$-dependent
current
\begin{equation}
    \begin{split}
        I\left(\delta\right) = \frac{e_0}{\pi\sqrt{m_{t}\left(\delta\right)}} \int d\epsilon \; &DOS_{2D}\left(\epsilon\right) f\left(\epsilon,\epsilon_{F,S}\right)
\\
        &\times \sqrt{\pm 2\left(\epsilon  - E_0 - \epsilon_{ToB}\right)}     \label{eq_current}
    \end{split}
\end{equation}
where $e_0$ is the elementary charge, $\epsilon_{ToB}$ the
$V_{gs}$-dependent shift of the ToB energy,
$f\left(\epsilon,\epsilon_{F,S}\right)$ the Fermi distribution
function, and $\epsilon_{F,S}$ the Fermi level in the source. From
Eq. \ref{eq_current} it can be inferred that the current $I_d$ should
linearly depend on $1/\sqrt{m_t\left(\delta\right)}$ because the DOS
is the same for all misalignment angles. This linear dependence can be
observed in Fig. \ref{fig:fig3}(b), where the ON-state currents from
Fig. \ref{fig:fig2}(b) are reported.

We then use Eq. \ref{dos_relation} and \ref{eq_current} to extract
$m_{DOS}$ and $m_{t}\left(\delta\right)$, following the procedure
outlined in \cite{Klinkert}. The results are compared to
$m_{t}\left(\delta\right)$, as calculated with  Eq. \ref{eq_mt}, in
Fig. \ref{fig:fig3}(c). An excellent agreement is demonstrated. Key
point is that  $m_{t}\left(\delta\right)$ exhibits a plateau for
$\delta \leq 20^{\circ}$, which reflect into an almost stable current
in this range. Hence, Figs. \ref{fig:fig3}(b) and (c) validate the form
of Eq. \ref{eq_current}.

The latter is finally recalled to compute the misalignment-induced
current reductions w.r.t. to the maximum current value for any 2D
material with an ellipsoidal bandstructure similar to {\color{black}
  BP}. A map of the current reductions as a function of the 
$m_{\Gamma-X}/m_{\Gamma-Y}$ ratio and the misalignment angle $\delta$
is presented in Fig. \ref{fig:fig3}(d) under the assumption of a fixed
DOS. It can be seen that for $m_{\Gamma-X}/m_{\Gamma-Y}<0.1$ the
ON-state current only marginally decreases up to $\delta\leq20^{\circ}$,
by $25\%$ if the misalignment is pushed to $40^{\circ}$. It is clear
that the magnitude of the ON-state current depends on the DOS of each
2D material. Nevertheless, a region where the current is almost
insensitive to the misalignment angle can be expected in all cases. 

\section{Conclusion}

The effect of orientation misalignment caused by bandstructure
anisotropies has been studied for $n$- and $p$-type black phosphorus
FETs. A reduction of the ON-state current below $30\%$ is found for
misalignment angles $\delta\leq 50^{\circ}$, while the performance
remains excellent for $\delta \leq 20^{\circ}$, {\color{black} with and
  without scattering}. The effective mass along the transport
direction explains this behavior, {\color{black} which does not display
  a strong dependence on electron/hole-phonon and charged impurity
  scattering}. For materials with a similar ellipsoidal bandstructure
as {\color{black} BP}, regardless of the $m_{\Gamma-X}/m_{\Gamma-Y}$
ratio, minor current degradations are foreseen for misalignment angles
$\delta$ up to $20^{\circ}$. 


\section{Acknowledgments}
This research was supported by the Swiss National Science Foundation
  (SNSF) under Grant No. $200021\_175479$ (ABIME) and under the NCCR
  MARVEL. We acknowledge CSCS for awarding us access to Piz Daint
  under project s876.

\bibliography{paperp4_colored}
\bibliographystyle{IEEEtran}

\end{document}